\documentclass[sigconf, nonacm, timestamp]{acmart}

\usepackage[most]{tcolorbox}

\usepackage{graphicx}
\usepackage{ifthen}
\usepackage{calc}
\newlength{\mylength}
\newlength{\mylengthfc} 
\AtBeginDocument{%
  \providecommand\BibTeX{{%
    \normalfont B\kern-0.5em{\scshape i\kern-0.25em b}\kern-0.8em\TeX}}}

\setcopyright{acmcopyright}
\copyrightyear{2018}
\acmYear{2018}
\acmDOI{10.1145/1122445.1122456}

\acmConference[Woodstock '18]{Woodstock '18: ACM Symposium on Neural
  Gaze Detection}{June 03--05, 2018}{Woodstock, NY}
\acmBooktitle{Woodstock '18: ACM Symposium on Neural Gaze Detection,
  June 03--05, 2018, Woodstock, NY}
\acmPrice{15.00}
\acmISBN{978-1-4503-XXXX-X/18/06}

\newboolean{LONG}
\setboolean{LONG}{true} 

\begin{teaserfigure}
    \begin{tcolorbox}[colback=red!5!white, colframe=red!75!black]
This is an early and extended version of a later publication at the \textit{15th International Symposium on Foundations and Practice of Security}, FPS 2022, Ottawa, ON, Canada. See \url{https://doi.org/10.1007/978-3-031-30122-3_7} or project site at \url{https://camm.h-da.io}
\end{tcolorbox}
\end{teaserfigure}

\begin{document}

\title[Crypto-agility Maturity Model]{Towards a maturity model for crypto-agility assessment}

\author{Julian Steffen Hohm}
\affiliation{%
  \institution{Hochschule Darmstadt}
  \country{Germany}
}

\author{Andreas Heinemann}
\affiliation{%
  \institution{Hochschule Darmstadt}
  \country{Germany}
}

\author{Alexander Wiesmaier}
\affiliation{%
  \institution{Hochschule Darmstadt}
  \country{Germany}
}

\renewcommand{\shortauthors}{Hohm, et al.}

\newcommand{\todo}[1]{\textcolor{red}{#1}}
\newcommand{\ca}{crypto-agility} 
\newcommand{\calong}{cryptographic agility}

\newcommand{\camm}{CAMM} 
\newcommand{\cammlong}{Crypto-Agility Maturity Model}
\newcommand{\cammsmalllong}{crypto-agility maturity model}
\newcommand{\mm}{maturity model}

\newcommand{\levelZERO}{\textsc{Initial / Not possible}}
\newcommand{\levelONE}{\textsc{Possible}}
\newcommand{\levelTWO}{\textsc{Prepared}}
\newcommand{\levelTHREE}{\textsc{Practiced}}
\newcommand{\levelFOUR}{\textsc{Sophisticated}}

\newcommand{\out}{Outcome}

\begin{abstract}
This work proposes the \cammlong{} (\camm{} for short), a
maturity model for determining the state of \ca{} of a given software or IT landscape.  \camm{} consists of five levels, for each level a set of requirements have been formulated based on literature review. Initial feedback from field experts confirms that \camm{} has a well-designed structure and is easy to comprehend. Based on our model, the \calong{} of an IT landscape can be systematically measured and improved step by step. We expect that this will enable companies and to respond better and faster to threats resulting from broken cryptographic schemes. This work serves to promote \camm{} and encourage others to apply it in practice and develop it jointly.
\end{abstract}


\maketitle

\section{Introduction}
\label{sec:intro}

In the light of NIST's current initiative to standardize post-quantum cryptographic (PQC) algorithms \cite{computer_security_division_nist_2017} in order to withstand potential attacks by powerful quantum computers, for example by running Shor's algorithm \cite{shor_algorithms_1994} against RSA, the more fundamental concept of \textit{\calong{}} (\textit{\ca{}} for short) has received an increasing focus recently, at least as a desirable property in the  context of PQC issues \cite{Niederhagen.2017,OttidentifyPQCchallenges,MehrezProperties, NAP24636, grote2019paradigm, erbacher2018crypto}. Although there is no common understanding of \ca{} in general, it is often associated with the ability to replace a cryptographic scheme in an agile manner with very little efforts. 

Following the view of Ott et. al \cite{OttidentifyPQCchallenges}, in our opinion, \ca{} needs to be discussed and addressed in a broader sense. Ott et. al propose the concept of \textit{modalities} for an expanded notion of \ca{}. For example, context agility refers to a crypto-agile solution, where cryptographic algorithms and strength policies have the flexibility to be derived automatically from system attributes.

In addition, the demand for \ca{} can be seen independently of the current PQC standardization activities of NIST. Just image, NIST will standardize crypto-scheme \textit{X} in the near future and \textit{X} is broken by an even more clever algorithm than Shor's algorithm in a few years ahead. This will put us back in the same situation again and we need to replace \textit{X} by a stronger scheme \textit{Y} in no time. In the face of the quantum computer threat, we should now transform our crypto solutions into fully comprehensive crypto-agile solutions.

A first step in this direction is to assess the current \ca{} of a particular software or IT system. With this knowledge, further development towards a \ca{} solution can then take place. In this work, we propose the \textit{\cammlong{}}{} (\camm{}) to determine the \ca{} level of a given software or IT System. \camm{} is composed of 5 maturity levels. For a system to reach a certain level, a number of given requirements must be met. We have formulated these requirements based on an intensive literature review on identified \ca{} publications and assigned them to the appropriate levels. With \camm{} at hand, IT managers can systematically assess their IT infrastructure and derive concrete measures to further develop their IT landscape in the direction of \ca{}.

The further text is structured as follows. 
Section \ref{sec:requirements} identifies important requirements, aspects and properties of \ca{} derived from literature, which we will later integrate into our \mm{}. This is followed by the methodology used in order to develop our \mm{} for \ca{} (Section \ref{sec:methodology}). The model itself is described in Section \ref{sec:camm}. We have set up an accompanying website at
\ifthenelse{\boolean{LONG}} 
{ 
\url{https://camm.h-da.io}
}
{ 
\textit{blinded for review}
}
in order to disseminate the model more widely. A brief preliminary evaluation of \camm{} is provided in Section \ref{sec:eval}, followed by a short discussion and outlook in Section \ref{sec:summary}. There we point out issues we would like to address in the future.

\section{\calong{}: Definitions, requirements and aspects}
\label{sec:requirements}

To the best of our knowledge, the notion of \calong{} was first mentioned around 2009/2010 by Bryan Sullivan \cite{microsoftagility2009, sullivanblackhat} as a programming style for abstracting .NET code from hard-coded use of a concrete hash algorithm, in his case MD5.
The term was also coined in 2011 in RFC 6421 \cite{nelsonRadiusRFC6421} as a communication protocol property. Since then, several authors have used the term in different manners. Without claiming completeness, the following understandings can be found in literature. 
According to McKay at NIST \cite{NAP24636} \ca{} includes (1) \emph{the ability for machines to select their security algorithms in real time and based on their combined security functions;} (2) \emph{the ability to add new cryptographic features or algorithms to existing hardware or software, resulting in new, stronger security features;} and (3) \emph{the ability to gracefully retire cryptographic systems that have become either vulnerable or obsolete.} Mehrez and el Omri \cite{MehrezProperties} and Schneider \cite{NAP24636} stress how easy the migration from one crypto scheme to another can take place. Schneider adds the aspect of remaining interoperable after a certain hard- or software has evolved. More recently, \calong{} has been mentioned in the context of PQC migration tasks \cite{review, RichteragileQC,Alnahawi21migration}. 

In addition to the above understandings, at least the following requirements and aspects for \calong{} were requested by different authors:
\textit{IDs} (for algorithms or sets of algorithms),
\textit{transitioning},
\textit{key management},
\textit{interoperability (mandatory algorithms)},
\textit{balancing security strengths},
\textit{opportunistic security},
\textit{(effective) migration mechanism}
\cite{housleyRFC7696}.
\textit{Measurability},
\textit{interpretability},
\textit{enforceability},
\textit{security},
\textit{performance}
\cite{OttidentifyPQCchallenges}.
\textit{Switch between crypto schemes in realtime},
\textit{support for heterogenous environments},
\textit{policy-aware access to crypto primitives}, 
\textit{automatability (centralized)}, 
\textit{scalability} 
\cite{Infosec}.
\textit{Extensibility},
\textit{removeability},
\textit{interoperability},
\textit{flexibility},
\textit{fungibility},
\textit{reversability},
\textit{updateability},
\textit{transition mechanism},
\textit{backwards compatibility}
\cite{MehrezProperties}.
\textit{Testable} \cite{cryptosense}, 
\textit{usage of SDKs, crypto APIs} \cite{utimaco, Niederhagen.2017}, 
\textit{preparing for failure} \cite{NAP24636}  

These requirements and aspects vary in granularity and are sometimes vague in their meaning and implementation. 
Still, we managed to organize these requirements and address them at the different levels of \camm{}. 
\ifthenelse{\boolean{LONG}} 
{ 
Appendix \ref{appendix:requirements} summarizes the requirements and provides the sources for each requirement. The tables can also be found at \url{https://camm.h-da.io}.
}
{ 
The accompanying website at \textit{blinded for review} lists the requirements and their source in a compact and tabular format.
}

\section{Methodology}
\label{sec:methodology}

We  describe which methodology was used in order to develop \camm{}. In general, there are two types of maturity models. First \textit{focus area maturity models} \cite{FocusArea} such as the Dynamic Architecture Maturity Matrix (DyAMM), the Test Process Improvement (TPI) model and the Software Product Management (SPM) maturity matrix \cite{van2007building,tpi,spm}. Second, \textit{fixed-level/staged maturity models} \cite{becker_developing_2009,MMinReserarch} such as CMM resp. CMMI \cite{CMM, CMMI}. We decided to aim at a fixed-level model because these type of models are far more widely used, easier to comprehend, and allow for more concrete and comparable statements \cite{FocusArea, steenbergen2007}. In addition, focus area maturity models need more data and more experience in the \ca{} domain \cite{FocusArea} and seemed inappropriate to us, as \ca{} research is still in its infancy.

\begin{figure}[htbp]
  \centering
  \includegraphics[scale=0.70]{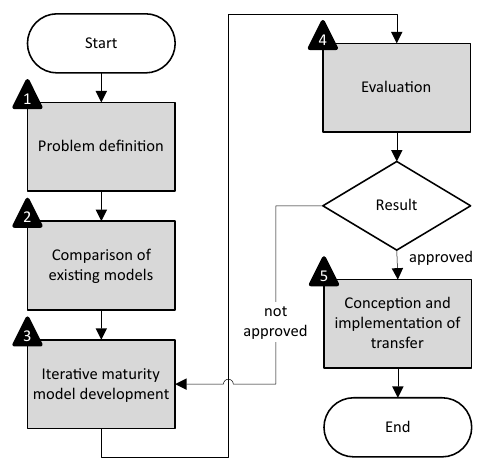}
  \caption{Applied procedure model for developing CAMM, adapted from \cite{becker_developing_2009}}
  \label{fig:beckerprocess}
\end{figure}

Lasrado et. al \cite{lasrado2015maturity} identified three meta models \cite{becker_developing_2009,meta39,Bruin.}, all describing a step by step iterative sequential approach for developing a maturity model. For \camm{}, we followed Becker et al. \cite{becker_developing_2009} as we judge this approach as the most comprehensive and detailed methodology for the development of fixed-level models. We, therefore, used it as a starting point and adapted it for this work. 
As a result, the following top-down approach was applied which consists of five phases/steps. We briefly describe each step and our findings on the basis of Fig. \ref{fig:beckerprocess}.

\subsubsection*{Step 1: Problem definition}

The first step is to clarify whether and why the lack of a maturity model is currently a problem for \ca{} and therefore whether it is worthwhile to develop a \cammsmalllong{}. 

\subsubsection*{\out{}:} A suitable \cammsmalllong{} can help to further expand the state of knowledge on \ca{}, as requested by several authors \cite{OttidentifyPQCchallenges, erbacher2018crypto, Alnahawi22CryptoAgility}.
A suitable maturity model will therefore allow us to better understand the state of \ca{} of a given IT landscape and thus, for example, be better prepared for the emerging threat of quantum computing.

\subsubsection*{Step 2: Comparison of existing models}
In order to solve the previously defined problem, it is researched whether existing models can be used, improved or adapted in the given context or if an entirely new model has to be developed.

\subsubsection*{\out{}:}

Numerous maturity models have already been identified in the literature \cite{MMinReserarch, TobiasMettler.2009}. The most popular CMMI is focused on processes and is thus not applicable for describing the maturity of a system's \ca{}. Models more specialized in security like the Cybersecurity Capability Maturity Model (C2M2) \cite{C2M2}, the Cybersecurity Maturity Model Certification (CMMC) \cite{CMMC}, the Systems Security Engineering — Capability Maturity Model (SSE-CMM) (ISO/IEC 21827:2008) \cite{ISO21827} and the Information Security Management Maturity Model (O-ISM3) \cite{O-ISM3} are mainly focused on improving processes and practices for cybersecurity and consider \ca{} only briefly, if at all. Classification in the Post-Quantum Cryptography (PQC) Maturity Model \cite{digicert} only focuses on measures against quantum computing threats. As previously stated, an IT landscape should be as crypto-agile as possible in order to be prepared for future threats, therefore a more general \cammsmalllong{} is needed. The Crypto Agility Risk Assessment Framework (CARAF) only provides a means to analyze and evaluate the risk that results from the lack of \ca{}  \cite{CARAF2021}. Cryptosense reduce their Maturity Model \ca{} to technologies and practices provided by their proprietary Cryptography Lifecycle Management (CLM) \cite{cryptosenseCLMMM}. To sum-up, we have not been able to find an appropriate \mm{} that supports all the requirements formulated for \ca{} (see Section \ref{sec:requirements}) and therefore developed \camm{} from scratch. 


\subsubsection*{Step 3: Iterative maturity model development} 

The main phase of the process is used to condense the existing \ca{} literature requirements into specific dimensions. 
First, we determine the objectives and development areas of the iteration. These are the basis for a first version of
a model. Via an iterative process the model is evaluated and evolves until all requirements fit into the various maturity levels and no inconsistencies exist.

\subsubsection*{\out{}:} 
The results are our five maturity levels and the corresponding requirements leading to \camm{} (see Section \ref{sec:camm}).

\subsubsection*{Step 4: Evaluation} 

After the development of a \mm{}, measures for evaluation, publication, and maintenance of the model are planned. In contrast to the process model according to \cite{becker_developing_2009}, the step of evaluation is preceding the publication of the model in order to be able to implement potential improvements identified therein before gaining first publicity. 

\subsubsection*{\out{}:} Our preliminary results show good structure and comprehensibility for \camm{} (see Section \ref{sec:eval}). 

\subsubsection*{Step 5: Conception of transfer and implementation} 

Becker et al. \cite{becker_developing_2009} suggest publishing the \mm{} and, in addition, establishing a website where the maturity level can be determined. In addition, the website should provide for a survey of model acceptance, which may indicate the need for further development.

\subsubsection*{\out{}:}

Following Becker et al. we set up a website at 
\ifthenelse{\boolean{LONG}} 
{ 
\url{https://camm.h-da.io}. 
}
{ 
\textit{blinded for review}.
}
This page is intended to serve as a central reference source for \camm{}. In addition to a brief description of the model, it contains the list of all requirements, literature references and publications. We will also provide a toolbox (new tools or refer to existing ones) to facilitate checking certain requirements in practice. CAMM is currently labeled with \textit{Version 1}. Further developments can then be tagged with higher version numbers. Additionally, the description of case studies that use \camm{} in practice is planned.

\section{The \cammlong{} }
\label{sec:camm}

The resulting \cammsmalllong{} represents our understanding of the current state of research on \ca{}. 
This model makes it possible to evaluate a given system according to its ability to implement  \ca{} requirements. At the same time, it forms a reference that encompasses the findings from both academia and practice and combines them into an understanding of the requirements of crypto-agile systems. 
\camm{}'s key components are the maturity levels and the requirements they contain at each level, along with their respective properties.

Since it is desirable to have the simplest possible classification while maintaining a high degree of precision, five maturity levels have emerged as an appropriate classification size for this model. This number also became a frequent standard value for maturity models due to the widespread adoption of the CMM and CMMI \cite{becker_developing_2009, lahrmann_inductive_2011}. 

\begin{table}
  \caption{CAMM maturity levels}
  \label{tab:camm stages}
  \begin{tabular}{cl}
    \textbf{Level}  & \textbf{Name}   \\
    \midrule
    0               & \levelZERO{} \\
    1               & \levelONE{} \\
    2               & \levelTWO{} \\
    3               & \levelTHREE{} \\
    4               & \levelFOUR{} \\
\end{tabular}
\end{table}

Table \ref{tab:camm stages} illustrates the five defined maturity levels. They are each described by their name and number and represent a certain development level of a system. Each level contains a certain number of requirements, all of which must be met in order to reach that level.  In other words, as soon as a requirement of a given level $X$ is violated, one falls back to level $X-1$ with level 0 as the lowest possible level.

To make the individual maturity levels more tangible, these and the motivations behind are be presented now. The requirements defined for a certain level are subsequently specified in Section \ref{sec:mmreqs}.

\subsubsection*{\levelZERO{}} 
This maturity level expresses the lowest maturity and is reached by default. The numbering explicitly begins at zero to illustrate the lack of evaluation or exclusion of \ca{}. This maturity level implies that there is at least one system or component that violates the requirements defined at level 1. Possible reasons for this may include hardware or software limitations that do not allow subsequent changes to the original design. Examples include devices in the IoT domain that are no longer supported by their manufacturer and platforms where cryptography is hard-coded to hardware features \cite{OttidentifyPQCchallenges}. Another familiar scenario is embedded systems that are inaccessible or where there is no vendor interest in enabling costly updates. 

\subsubsection*{\levelONE{}} 
This maturity level is reached by all systems that can be adapted so that their cryptography can respond dynamically to future crypto challenges. However, this does not require any specific activities to be carried out yet, but only the necessary primary conditions to be met. Many of the requirements are
typical for high-quality software or hardware design.
    
\subsubsection*{\levelTWO{}} 
Systems that already implement certain measures for \ca{}, but are not yet fully ready to actively realize it, are assigned to this level. The actual change of cryptographic functions still requires some preparatory work here and means a particular effort, but \ca{} is already seen as an implementable goal. 

\subsubsection*{\levelTHREE{}} 
Crypto-agility is practiced, i.e,  migration between different cryptographic methods is demonstrably, effectively, and securely feasible. In this case, systems can be assigned to this maturity level. Therefore, several conditions must be met to ensure the necessary hardware and software requirements and migration mechanisms. 

\subsubsection*{\levelFOUR{}} 
The highest maturity level is attained by systems that implement advanced capabilities in terms of \ca{}. They are particularly characterized by the fact that compatibility is not limited to a specific system but is applied in a broader infrastructure. Higher sophistication and practical testing of the measures allow for a fast migration between different cryptography schemes. This level of maturity should be strived for particularly by libraries and frameworks that are intended to be used in the context of \ca{}.

\begin{figure}[htbp]
  \centering
  \includegraphics[scale=0.45]{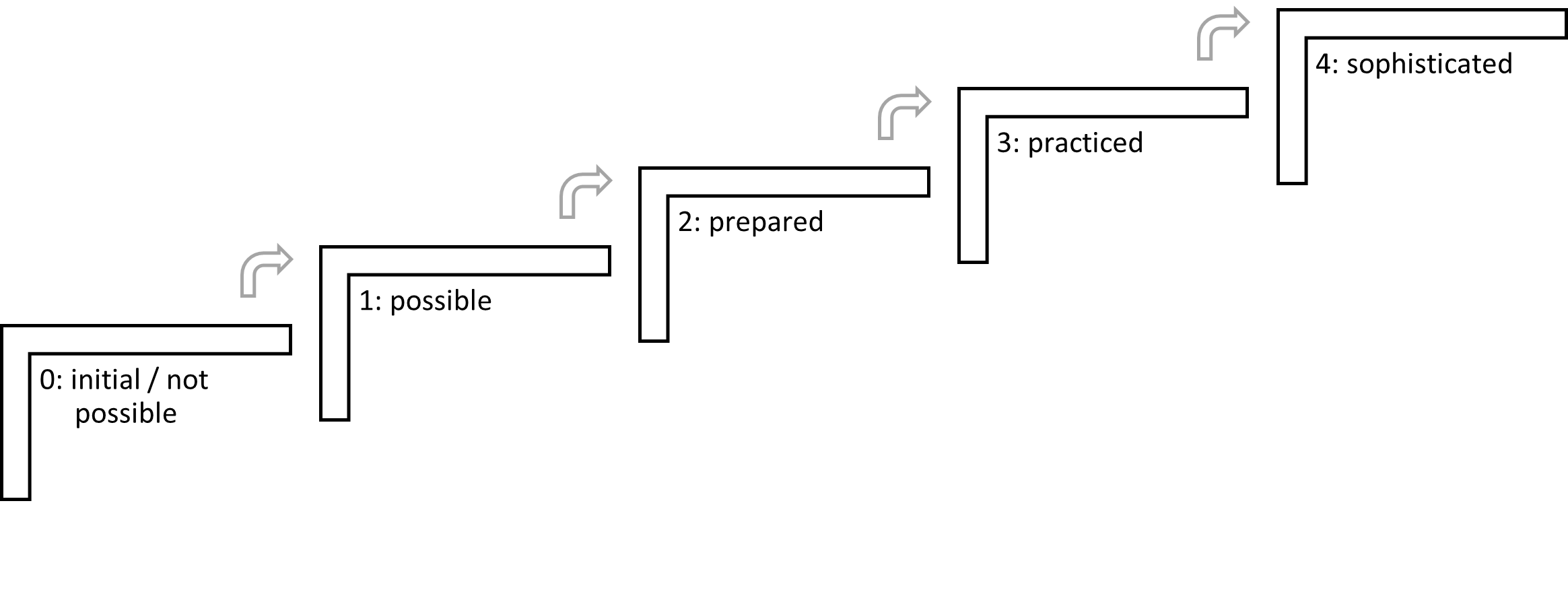}
  \caption{\camm{} -- Stage model perspective}
  \label{fig:CAMMstages}
\end{figure}

As shown in Fig. \ref{fig:CAMMstages} the defined maturity levels can be understood as a stage model. Maturity level 0 forms the lowest stage and symbolizes the lowest or no capability for \ca{}. The higher the maturity level, i.e., the higher the stage, the more agile a systems cryptography is. 


For each maturity level, different measures can be used to meet the requirements, but all must be suitable for achieving a certain level of maturity.

\subsection{Requirements}
\label{sec:mmreqs}

In the following, the requirements that are addressed at a certain level are presented. Please note, that there are no requirements associated with level 0 of \camm{}. Next, each requirement is assigned to a certain category. We have defined three different categories: \textit{Knowledge (K)}, \textit{Process (P)}, \textit{System property (S)}. This categorization allows the implementation of measures necessary to fulfill a certain requirement to be grouped and considered together. For example, requirements from the  category \textit{Knowledge} require a similar approach, e.g., the construction of a knowledge database.  

Each requirement is labeled by 'R' following a two-digit number. The first digit defines the corresponding maturity level, the second digit defines a sequential ID. This ID does not imply any order of priority. Example: R34 labels requirement 4 within level 3.

\subsubsection*{R10: System knowledge (K)} An adequate evaluation of the requirements of \ca{} demands detailed information about the corresponding system. System knowledge thus represents a requirement that must be met for a system to reach the first maturity level of \ca{}. Without detailed knowledge of the system and its domain a classification is not possible, and the \ca{} goal of measurability identified by \cite{OttidentifyPQCchallenges} cannot be met. Additionally, the ability to analyze the system is inevitable to assess the impact on a product or system of an intended change to one or more of its parts, or to diagnose a product for deficiencies or causes of failures, or to identify parts to be modified \cite{ISO25010}. 

\subsubsection*{R11: Updateability (P)} This property describes whether the system can be modified by those responsible and can be supplied with new software versions without the need to restrict its functionality \cite{ISO25010}. The underlying problem is the discovery of vulnerabilities in the system and its cryptography, which need to be fixed after detection. Like initially described by \cite{MehrezProperties} a system must therefore provide the ability to perform updates. A prerequisite for the development of such updates is system knowledge. 

\subsubsection*{R12: Extensibility (S)} Extensibility can be understood as the process of introducing new cryptographic algorithms and parameters. In the case of a new attack vector affecting previous algorithms, new, more secure alternatives can thus be retrofitted. The ability to add further cryptography methods, variants and protocols to a system can be understood as an acceptance criterion. This requirement originates from the \ca{} properties of \cite{MehrezProperties} and requires updateability of the software. 

\subsubsection*{R13: Reversibility (P)} In addition to updates, the reversibility of systems is also an important factor for \ca{}. The goal of this requirement is to allow a system to be reset to a previous state. However, unlike the therefore required updateability, this is not a property of the system but requires an organizational process. As described by \cite{MehrezProperties} a system needs to be able to return to previous version if the update does not operate as expected.

\subsubsection*{R14: Cryptography inventory (K)} A cryptography inventory is the documentation of the cryptography used in a system. In addition to simply documenting the algorithms and parameters used, it is also important to understand their level of security. Thus, this requirement is an aspect that presupposes a certain knowledge about the system and the current state of cryptographic security. The rationale for this requirement is the need to assess a system's security. If vulnerabilities in certain cryptographic implementations are discovered, \cite{erbacher2018crypto,cryptosense}, among others, demand a cryptography inventory to enable a fast evaluation of whether a system is affected by them, so countermeasures can be initiated if necessary. 

\subsubsection*{R20: Cryptographic modularity (S)} In the context of \ca{}, cryptographic modularity is understood as a system design that enables changes to the cryptographic components without affecting the functionality of the other system components. In the event of a vulnerability, the implementation of cryptographic functions, their parameters and primitives, can be replaced without affecting the system logic. System knowledge is required to evaluate this property. The necessity of this property is already confirmed by a large number of experts in this regard \cite{MehrezProperties,RichteragileQC,erbacher2018crypto,housleyRFC7696,Infosec}.
    
\subsubsection*{R21: Algorithm IDs (K)} Cryptographic procedures cannot be negotiated between two subsystems without a convention denoting algorithms and parameters. Thus, common knowledge is essential to uniquely associate algorithms with their specific parameters for building agile encrypted communication channels. For IETF protocols, a mechanism for identification is already mandatory and \cite{housleyRFC7696} also lists it as a requirement for \ca{}. 
    
\subsubsection*{R22: Algorithm intersection (S)} This requirement implies that all subsystems support a common set of cryptographic algorithms. Without this intersection of algorithms, secure communication cannot be established and therefore, at least one algorithm must be defined that all subsystems mandatorily support. This principle is also prescribed for IETF protocols and highlighted in RFC 7696 \cite{housleyRFC7696}  for cryptographic algorithms agility.

\subsubsection*{R23: Algorithm exclusion (S)}
As a counterpart to mandatory algorithms, this property requires a way to exclude supported algorithms from use. The reason for this is that algorithms known to be vulnerable should not be used to maintain the security of the system. However, as originally stated in  \cite{housleyRFC7696,MehrezProperties}, the exclusion of algorithms simultaneously affects the interoperability of (sub)systems. 

\subsubsection*{R24: Opportunistic security (S)}
This requirement pursues the goal of always using the strongest algorithm supported by the respective communication partners and that is appropriate in the respective application context. If a particular algorithm cannot be used for some reason, it is better to use a less secure one than to communicate without cryptography altogether \cite{rfc7435opportunistc}. Even with weak cryptography, resilience to attacks is increased, albeit minimally in the worst case. This property requires an intersection of multiple algorithms, assignable by their ID, which are ranked by their security in the crypto inventory. However, this property may be constrained by policies that, for example, mandate a certain level of security. This concept is described in \cite{rfc7435opportunistc} and proposed by \cite{housleyRFC7696} for the \ca{} of IETF protocols. 

\subsubsection*{R30: Policies (P)}
Policies serve to constrain the algorithms that are allowed and their parameters \cite{Infosec}. Without them, insecure algorithms or algorithms inappropriate for the context could be used. As an essential factor for the security of systems, they are an essential step in the system design process and ongoing management. Their specification can constrain the algorithm intersection, result in algorithm exclusion and hinder opportunistic security. 

\subsubsection*{R31: Performance awareness (K)}
Performance awareness describes that the additional overhead for \ca{} is known and accepted \cite{OttidentifyPQCchallenges,housleyRFC7696}. If the additional effort of adaptations and deployment for \ca{} is not known and accounted for, it may result in unexpected performance degradation and cost, which could cause the system to fail to meet specific requirements. To understand the impact of \ca{} on the specific circumstances of the system, a deep knowledge of the system is essential.

\subsubsection*{R32: Hardware modularity (S)}
In addition to the cryptographic modularity, a loose binding between hardware and software is another critical property to support \ca{} \cite{MehrezProperties,OttidentifyPQCchallenges,RichteragileQC}. It represents the possibility of further additions or replacements to both hardware and software independently and compatible with each other. 

\subsubsection*{R33: Testing (P)}
 If requirements are not tested, no conclusions can be drawn about their compliance and the quality of the system. A crucial part of the process towards \ca{} is regularly testing the system for compliance with crypto-agile requirements \cite{cryptosense,OttidentifyPQCchallenges,acc2021race}. Therefore, test criteria must exist to determine which properties the system must meet. The \ca{} requirements defined here should be part of the testing.

\subsubsection*{R34: Enforceability (P)}
The appropriate measures to achieve \ca{} must be taken for all system areas. When the required techniques are carefully specified, a crypto-agile design can be effectively mandated and enforced \cite{OttidentifyPQCchallenges}. If the \ca{} requirements and techniques are not known, no process can be initiated to implement them. This work contributes to increasing the enforceability of \ca{} by providing the necessary requirements from which actions for systems can be derived and subsequently implemented.

\subsubsection*{R35: Security (K)}
Maintaining security is another integral factor when implementing \ca{}\cite{OttidentifyPQCchallenges}. This requires ensuring that the system is not vulnerable to the threats of various attacks. If this and the maintenance of the overall security objectives cannot be ensured by the particular implementation of the cryptographic techniques, they do not fulfill the expected benefits. In order to obtain knowledge about the security state of a given system at this level, regular checks must confirm a healthy security state of the used cryptography and transition mechanisms.

\subsubsection*{R36: Backwards compatibility (S)}
This property refers to the time-limited phase during which new versions are compatible with older states of the system \cite{MehrezProperties}. This transition period is necessary because the overall system will not be functional if different concurrently used version states of the subsystems cannot interact. To ensure that the system does not remain permanently vulnerable to attack due to security vulnerabilities in old versions, backward compatibility must only exist within a limited period.

\subsubsection*{R37: Transition mechanism (P)}
The process with which it is possible to migrate between cryptographic methods is mainly responsible for the dynamics of \ca{}. Here the transition mechanism constitutes the strategy that ensures the operability of the overall system for the transition period of performing an update \cite{MehrezProperties,housleyRFC7696,cryptosense,erbacher2018crypto}. Without a regulated and secure process, the compatible switch to the new version is not possible without failures. Therefore, in crypto-agile systems, a methodology must exist that regulates compatibility and secure communication between subsystems in different version states for the time-limited duration of the transition and results in all subsystems being at the new version state afterwards. 

\subsubsection*{R38: Effectiveness (P)}
The effectiveness of \ca{} expresses that the process of migrating between cryptographic algorithms must be feasible in a reasonable amount of time \cite{OttidentifyPQCchallenges}. The rationale behind this requirement is to ensure security. If the migration takes longer than the security of the algorithms can be guaranteed, the whole system is vulnerable. 

\subsubsection*{R40: Automation (P)}
This is a more advanced measure of \ca{}, used primarily when a wide range of platforms should be made crypto-agile. Through automation, customizations to crypto-agile modules require no, or very little user interaction and must be able to be performed automatically based on predefined attributes such as time and context, without interaction with the system \cite{Infosec}. Thus, the effort for processes within \ca{} is minimized by reducing the manual interactions required for it to a minimum. 

\subsubsection*{R41: Context independence (S)}
This property requires that requirements and techniques are portable such that they can be used for other cryptographic scenarios. Also known as interpretability and flexibility \cite{OttidentifyPQCchallenges,MehrezProperties}.
This property circumvents the problem that the approaches used for \ca{} are each applicable only in their specific context, and thus a separate solution must be found for each domain. 

\subsubsection*{R42: Scalability (P)}
To avoid repeating the process of developing a system module for \ca{} for all systems, the crypto module must be deployable on other systems without substantial adjustments. Otherwise, the implementation would always be tied to specific platforms, and each additional system to be made crypto-agile would again require a considerable development effort \cite{Infosec}. 

\subsubsection*{R43: Real-time (P)}
Another challenge for crypto agility processes is the time required to run them. In this context, real-time systems assure that adaptations to cryptographic functions become active in the production system within a defined period. The goal here is to ensure that adaptations to the cryptography do not require re-coding and re-booting of the system \cite{Infosec}.

\subsubsection*{R44: Interoperability (S)}
A necessary consideration for \ca{} is communication beyond the context of a single system or platform. Here the property of interoperability expresses that different crypto-agile systems are compatible with each other \cite{MehrezProperties,housleyRFC7696}. The challenge is that connected systems might not always be in one's own administrative domain and therefore cannot be influenced directly. Nevertheless, the global IT infrastructure should be compatible with all approaches to \ca{}. To achieve this, crypto-agile systems must enable information exchange with all foreseen communication partners based on the information in their respective specifications.

\ifthenelse{\boolean{LONG}} 
{ 
All requirements are listed in tabular form in Appendix \ref{appendix:requirements}. For each requirement, a brief description is given, the problem is outlined, an acceptance criterion is described, and dependencies and the literature source on which the criterion is based are given. 
}
{ 
All requirements are listed in tabular form on the accompanying website at \textit{blinded for review}. For each requirement, a brief description is given, the problem is outlined, an acceptance criterion is described, and dependencies and the literature source on which the criterion is based are given.
}

\section{Preliminary Evaluation} 
\label{sec:eval}

An initial evaluation of \camm{} was conducted as part of a 
\ifthenelse{\boolean{LONG}} 
{ 
master's thesis\cite{Hohm2021} 
}
{ 
\textit{blinded for review} study
}
together with a partner company. First, semi-structured interviews were conducted with two inhouse-experts (one Security Officer and one Consultant System Software Architect) with regard to the comprehensibility of \camm{}. 
In addition, the relevance of \ca{} and post-quantum cryptography within the company was assessed. While PQC is not yet seen as relevant, they acknowledge the benefit of the more general \calong{} concept.
Both experts confirmed the comprehensibility of \camm{} and appreciated the sound structure.
On the other hand, both assume that \camm{} must be brought into use so that more experience can be gained with the model. They assume that especially level 4 \levelFOUR{} may be be extended or modified.

\section{Outlook}
\label{sec:summary}

Based on an extensive literature review on \ca{} and the identified lack  of a corresponding maturity model, we propose \camm{} to fill this gap. Now, a broad application of \camm{} has to take place to verify the validity and usefulness of our model and if necessary to adjust the requirements on a respective level. In addition, a toolbox should be developed that helps to verify the individual requirements at a certain level with as little effort as possible.


\bibliographystyle{ACM-Reference-Format}
\bibliography{main}

\ifthenelse{\boolean{LONG}} 
{ 
    \appendix
    \section{Requirements}
\label{appendix:requirements}

See next page onwards.

\setlength{\mylength}{\textwidth-2.6cm}%
\setlength{\mylengthfc}{1.8cm}


\begin{table*}[h!]
  \caption{Requirements on Level 1}
  \label{tab:requirementl1}
  \begin{tabular}{p{\mylengthfc}p{\mylength}}
    \textbf{Name}                   & Systemknowledge   \\
    \midrule
    \textbf{ID}                     & 10 \\
    \textbf{Category}               & Knowledge   \\
    \textbf{Description}            & For \ca{} requirements to be effectively evaluated, detailed knowledge of the affected system and its environment is required.  \\
    \textbf{Problem}                & Without knowledge about the systems and understanding about their domain, no assertions can be made about them and \ca{} cannot be measured. \\
    \textbf{Acceptance}             &  An in-depth understanding of the structure and operation of the systems being evaluated is available.   \\
    \textbf{Dependency}             & -       \\
    \textbf{Source}                 &  \cite{OttidentifyPQCchallenges}  \\
    \textbf{Example}                & Access to source code and/or hardware specification. Black boxes cannot be evaluated.   \\       
\end{tabular}

\vspace*{2ex}

  \begin{tabular}{p{\mylengthfc}p{\mylength}}
    \textbf{Name}                   & Updateability   \\
    \midrule
    \textbf{ID}                     &    11 \\
    \textbf{Category}               &    Process \\
    \textbf{Description}            &    Maintainers can modify the system and provide updates to new software versions.  \\
    \textbf{Problem}                &    If vulnerabilities are identified in the system and its cryptography, it should be possible to fix them.\\
    \textbf{Acceptance}             &    Performing updates with modifications is possible. \\
    \textbf{Dependency}             &    10 \\
    \textbf{Source}                 &    \cite{MehrezProperties, medical} \\
    \textbf{Example}                &    Mobile apps are often modified by updates. Modifiability is not possible for legacy devices without support.\\
\end{tabular}

\vspace*{2ex}

  \begin{tabular}{p{\mylengthfc}p{\mylength}}
    \textbf{Name}                   & Extensibility   \\
    \midrule
    \textbf{ID}                     &    12 \\
    \textbf{Category}               &    System property\\ 
    \textbf{Description}            &    The system can be extended with new cryptographic algorithms and parameters.\\
    \textbf{Problem}                &    In the event of a new attack vector that affects all previous algorithms, new, secure alternatives need to be added.\\
    \textbf{Acceptance}             &    Further cryptography methods, variants and protocols can be added.\\
    \textbf{Dependency}             &    11\\
    \textbf{Source}                 &    \cite{MehrezProperties,NAP24636}\\
    \textbf{Example}                &    Additional capacity that may be needed later is considered in the design of a system. Not possible for IOT devices that are already at the memory limit or whose performance is not sufficient for more computationally intensive algorithms.      
\end{tabular}

\vspace*{2ex}

  \begin{tabular}{p{\mylengthfc}p{\mylength}}
    \textbf{Name}                   & Reversibility   \\
    \midrule
    \textbf{ID}                     &    13 \\
    \textbf{Category}               &    Process\\
    \textbf{Description}            &    The system can be rolled back to a previous state.\\
    \textbf{Problem}                &    If an update results in problems, the system can be can be rolled back to a previous, functional state.\\
    \textbf{Acceptance}             &    Rollbacks to previous versions are possible.\\
    \textbf{Dependency}             &    10\\
    \textbf{Source}                 &   \cite{MehrezProperties}\\
    \textbf{Example}                &    Due to a bug in a system update the system does not behave as expected and is rolled back to a previous state.
    \end{tabular}

\vspace*{2ex}

\end{table*}

\begin{table*}
  \caption{Requirements on Level 1 (continued)}

  \begin{tabular}{p{\mylengthfc}p{\mylength}}
    \textbf{Name}                   & Cryptography inventory   \\
    \midrule
    \textbf{ID}                     &    14 \\
    \textbf{Category}               &    Knowledge\\
    \textbf{Description}            &    The cryptographic functions used are documented and their current security level is known.\\
    \textbf{Problem}                &    In order to assess whether the system is affected by known vulnerabilities in certain cryptography variants, there must be an overview of the cryptography implementations used.\\
    \textbf{Acceptance}             &     A listing of the cryptographic methods used, their parameters and intended use is available, and current developments and recommendations for action on cyber security are observed.\\
    \textbf{Dependency}             &    10\\
    \textbf{Source}                 &    \cite{gartner, erbacher2018crypto}\\
    \textbf{Example}                &    Inventory as a table with table with the following information: cryptography methods, primitives used, key length, purpose of use, security level, date of deployment, date of deactivation. Trends and developments in cryptographic security are tracked at conferences and in related publications.     
\end{tabular}

\end{table*}

\newpage


\newpage

\begin{table*}
  \caption{Requirements on Level 2}
  \label{tab:requirementl2}
  \begin{tabular}{p{\mylengthfc}p{\mylength}}
    \textbf{Name}                   &    Cryptographic modularity\\
    \midrule
    \textbf{ID}                     &    20 \\
    \textbf{Category}               &    System property\\
    \textbf{Description}            &    The system is modularly designed in such a way that changes to the cryptographic components do not affect the functionality of the other system components coupled to it.\\
    \textbf{Problem}                &    If the implementation of the cryptographic functions contains vulnerabilities, it can be replaced by another one without affecting the rest of the system logic.\\
    \textbf{Acceptance}             &    The specific implementation of the cryptographic functions, their parameters and primitives can be exchanged without affecting the other system components.\\
    \textbf{Dependency}             &    10, 11\\
    \textbf{Source}                 &    \cite{housleyRFC7696,MehrezProperties,RichteragileQC,Infosec,erbacher2018crypto}\\
    \textbf{Example}                &    An API separates cryptographic functions and business logic.\\       
\end{tabular}

\vspace*{2ex}

  \begin{tabular}{p{\mylengthfc}p{\mylength}}
    \textbf{Name}                   &    Algorithm IDs\\
    \midrule
    \textbf{ID}                     &    21 \\
    \textbf{Category}               &    Knowledge\\
    \textbf{Description}            &    The algorithms used are uniquely identifiable.\\
    \textbf{Problem}                &    Without a common understanding of the conventions used to designate algorithms and the parameters used, no cryptographic procedures can be negotiated between two subsystems.\\
    \textbf{Acceptance}             &    Each algorithm used with its specific parameters is uniquely referenced by an identifier.\\
    \textbf{Dependency}             &    14\\
    \textbf{Source}                 &    \cite{housleyRFC7696}\\
    \textbf{Example}                &    IDs of the TLS ciphersuites\\       
\end{tabular}

\vspace*{2ex}

  \begin{tabular}{p{\mylengthfc}p{\mylength}}
    \textbf{Name}                   &    Algorithm intersection\\
    \midrule
    \textbf{ID}                     &    22 \\
    \textbf{Category}               &    System property\\
    \textbf{Description}            &    All subsystems share a common set of cryptographic algorithms.\\
    \textbf{Problem}                &    A common intersection of algorithms is required to enable interoperability of different sub-systems.\\
    \textbf{Acceptance}             &    At least one algorithm is defined that is supported by all subsystems.\\
    \textbf{Dependency}             &    21\\
    \textbf{Source}                 &    \cite{housleyRFC7696}\\
    \textbf{Example}                &    TLS 1.3 prescribes TLS\_AES\_128\_GCM\_SHA256\\       
\end{tabular}

\vspace*{2ex}

  \begin{tabular}{p{\mylengthfc}p{\mylength}}
    \textbf{Name}                   &    Algorithm exclusion\\
    \midrule
    \textbf{ID}                     &    23 \\
    \textbf{Category}               &    System property\\
    \textbf{Description}            &    The system can disable the use of supported algorithms.\\
    \textbf{Problem}                &    If certain algorithms are vulnerable, they should not further be used.\\
    \textbf{Acceptance}             &    If an algorithm is marked as obsolete, it is no longer used for cryptographic tasks.\\
    \textbf{Dependency}             &    11, 21, 44\\
    \textbf{Source}                 &    \cite{housleyRFC7696,MehrezProperties,NAP24636}\\
    \textbf{Example}                &    Legacy algorithms in TLS 1.3 may only be used to ensure backward compatibility with TLS 1.2. MD5 is still widely used for fast, manual checking of downloads for integrity; its use for security-related cryptographic purposes should be discouraged today due to the comparably low collision resistance and the resulting attack vectors.\\       
\end{tabular}

\end{table*}

\begin{table*}
  \caption{Requirements on Level 2 (continued)}

  \begin{tabular}{p{\mylengthfc}p{\mylength}}
    \textbf{Name}                   &    Opportunistic security\\
    \midrule
    \textbf{ID}                     &    24 \\
    \textbf{Category}               &    System property\\
    \textbf{Description}            &    The system always uses the strongest available algorithm.\\
    \textbf{Problem}                &    If a certain algorithm cannot be used for certain reasons, it is better to use the next most secure algorithm than to discard cryptography altogether.\\
    \textbf{Acceptance}             &    The choice of cryptographic algorithms is based on the greatest possible security that is supported by all the subsystems involved, while complying with possible guidelines.\\
    \textbf{Dependency}             &    11, 21, 22, 23, 30\\
    \textbf{Source}                 &    \cite{housleyRFC7696}\\
    \textbf{Example}                &    Despite weaknesses of MD5, it continues to be used when it is the only mutually supported and therefore the strongest option.\\       
\end{tabular}

\end{table*}


\newpage

\begin{table*}
  \caption{Requirements on Level 3}
  \label{tab:requirementl3}
  \begin{tabular}{p{\mylengthfc}p{\mylength}}
    \textbf{Name}                   &    Policies\\
    \midrule
    \textbf{ID}                     &    30 \\
    \textbf{Category}               &    Process \\
    \textbf{Description}            &    Policies are used to restrict the allowed algorithms and their parameters.\\
    \textbf{Problem}                &    Without guidelines, insecure algorithms or algorithms unsuitable for the context could be used.\\
    \textbf{Acceptance}             &    Guidelines that define the minimum requirements for crypto-agile systems are specified and complied with by the algorithms used.\\
    \textbf{Dependency}             &    22, 23, 24, 35\\
    \textbf{Source}                 &    \cite{Infosec}\\
    \textbf{Example}                &    Algorithms used must have at least 2048 bit security level. Only certified solutions solutions may be used.\\       
\end{tabular}

\vspace*{2ex}

  \begin{tabular}{p{\mylengthfc}p{\mylength}}
    \textbf{Name}                   &    Performance awareness\\
    \midrule
    \textbf{ID}                     &    31 \\
    \textbf{Category}               &    Knowledge \\
    \textbf{Description}            &    The additional effort and impact of \ca{} is known and accepted.\\
    \textbf{Problem}                &    If the additional effort due to crypto-agility customizations and deployment is not known and accounted for, it may result in unexpected performance degradation and costs, which could cause the system to fail to meet certain requirements.\\
    \textbf{Acceptance}             &    Efforts for implementation and efficiency correspond to previously established requirements.\\
    \textbf{Dependency}             &    10\\
    \textbf{Source}                 &    \cite{OttidentifyPQCchallenges,housleyRFC7696}\\
    \textbf{Example}                &    The agile platform implies 25\% overhead in connection setup compared to using classical cryptography and 100 man-days of development effort.\\       
\end{tabular}

\vspace*{2ex}

  \begin{tabular}{p{\mylengthfc}p{\mylength}}
    \textbf{Name}                   &    Hardware modularity\\
    \midrule
    \textbf{ID}                     &    32 \\
    \textbf{Category}               &    System property\\
    \textbf{Description}            &    Hardware and software can be improved or exchanged independently of each other in a compatible manner.\\
    \textbf{Problem}                &    Agility is prevented if the system is restricted by the software or hardware used.\\
    \textbf{Acceptance}             &    Adaptations to the hardware are possible without affecting its compatibility with the software and vice versa.\\
    \textbf{Dependency}             &    11\\
    \textbf{Source}                 &    \cite{MehrezProperties,OttidentifyPQCchallenges,RichteragileQC,medical}\\
    \textbf{Example}                &    The system provides interfaces that enable the use of more powerful hardware, hardware for quantum cryptography, or more efficient implementation.\\       
\end{tabular}

\vspace*{2ex}

  \begin{tabular}{p{\mylengthfc}p{\mylength}}
    \textbf{Name}                   &    Testing\\
    \midrule
    \textbf{ID}                     &    33 \\
    \textbf{Category}               &    Process\\
    \textbf{Description}            &    The system is regularly tested for compliance with \ca{} requirements.\\
    \textbf{Problem}                &    Appropriate tests are required to measure the fulfillment of the requirements and quality of the system.\\
    \textbf{Acceptance}             &    There are test criteria that the system must meet, which are regularly checked for compliance.\\
    \textbf{Dependency}             &    10\\
    \textbf{Source}                 &    \cite{cryptosense,OttidentifyPQCchallenges,acc2021race}\\
    \textbf{Example}                &    Fulfillment of requirements gets verified regularly, Migration is run on a parallel test system.\\       
\end{tabular}

\end{table*}

\begin{table*}
  \caption{Requirements on Level 3 (continued)}
  \label{tab:requirementl3c}

  \begin{tabular}{p{\mylengthfc}p{\mylength}}
    \textbf{Name}                   &    Enforceability\\
    \midrule
    \textbf{ID}                     &    34 \\
    \textbf{Category}               &    Process\\
    \textbf{Description}            &    Crypto agility can be made mandatory for certain contexts.\\
    \textbf{Problem}                &    If the requirements and techniques of \ca{} are not known, it cannot be implemented.\\
    \textbf{Acceptance}             &    The required techniques are carefully specified and a crypto agile design can be effectively prescribed and enforced.\\
    \textbf{Dependency}             &    10, 30\\
    \textbf{Source}                 &    \cite{OttidentifyPQCchallenges}\\
    \textbf{Example}                &    This model provides a framework against which requirements and measures for a systems \ca{} can be developed and prescribed.\\       
\end{tabular}

\vspace*{2ex}

  \begin{tabular}{p{\mylengthfc}p{\mylength}}
    \textbf{Name}                   &    Security\\
    \midrule
    \textbf{ID}                     &    35 \\
    \textbf{Category}               &    Knowledge \\
    \textbf{Description}            &    The \ca{} mechanism is secure against attacks.\\
    \textbf{Problem}                &    Crypto agility is useless if it is vulnerable to attacks.\\
    \textbf{Acceptance}             &    No vulnerabilities are detected during audits of the system.\\
    \textbf{Dependency}             &    23, 30, 33\\
    \textbf{Source}                 &    \cite{OttidentifyPQCchallenges}\\
    \textbf{Example}                &    Regular audits confirm the security of the system.\\       
\end{tabular}

\vspace*{2ex}

  \begin{tabular}{p{\mylengthfc}p{\mylength}}
    \textbf{Name}                   &    Backwards compatibility\\
    \midrule
    \textbf{ID}                     &    36 \\
    \textbf{Category}               &    System property\\
    \textbf{Description}            &    New versions are compatible with older states of the system.\\
    \textbf{Problem}                &    The system as a whole is not functional if subsystems with different versions cannot interact with each other.\\
    \textbf{Acceptance}             &    New versions of the system must support all functions of the previous version during the transition phase.\\
    \textbf{Dependency}             &    11,22,37\\
    \textbf{Source}                 &    \cite{MehrezProperties}\\
    \textbf{Example}                &    Version 2 of a system ensures full functionality in conjunction with subsystems that are still in version 1.\\       
\end{tabular}

\vspace*{2ex}

  \begin{tabular}{p{\mylengthfc}p{\mylength}}
    \textbf{Name}                   &    Transition mechanism\\
    \midrule
    \textbf{ID}                     &    37 \\
    \textbf{Category}               &    Process\\
    \textbf{Description}            &    A strategy that ensures the functionality of the overall system for the transition period of performing an update.\\
    \textbf{Problem}                &    Agility requires transitions between old and new versions, which must follow a regulated and secure process.\\
    \textbf{Acceptance}             &    There is a methodology that regulates compatibility and secure communication between subsystems in different versions for the limited duration of the transition, resulting in all subsystems being at the new version status afterwards.\\
    \textbf{Dependency}             &    11, 21, 22, 23, 36\\
    \textbf{Source}                 &    \cite{MehrezProperties,housleyRFC7696,cryptosense,erbacher2018crypto}\\
    \textbf{Example}                &    All subsystems in version 1 are upgraded to version 2 without disruptions after a hybrid transition phase.\\       
\end{tabular}

\vspace*{2ex}

  \begin{tabular}{p{\mylengthfc}p{\mylength}}
    \textbf{Name}                   &    Effectiveness\\
    \midrule
    \textbf{ID}                     &    38 \\
    \textbf{Category}               &    Process\\
    \textbf{Description}            &    Migration between cryptographic algorithms must be feasible in a reasonable amount of time.\\
    \textbf{Problem}                &    If the migration takes longer than the security of the algorithms can be guaranteed, the entire system is vulnerable.\\
    \textbf{Acceptance}             &    After a vulnerability in the used cryptography becomes known, it is eliminated promptly and, if possible, before the attacks become practically relevant by migrating to a secure alternative.\\
    \textbf{Dependency}             &    14, 33, 37\\
    \textbf{Source}                 &    \cite{OttidentifyPQCchallenges}\\
    \textbf{Example}                &    x + y < z \cite{Mosca_xyz}\\       
\end{tabular}

\end{table*}


\begin{table*}
  \caption{Requirements on Level 4}
  \label{tab:requirementl4}
  \begin{tabular}{p{\mylengthfc}p{\mylength}}
    \textbf{Name}                   &    Automation\\
    \midrule
    \textbf{ID}                     &    40 \\
    \textbf{Category}               &    Process\\
    \textbf{Description}            &    Modifications to crypto-agile modules do not require manual interaction.\\
    \textbf{Problem}                &    \ca{} should require as little user interaction as possible.\\
    \textbf{Acceptance}             &    Changes and system components responsible for \ca{} can be made based on pre-determined attributes, such as time and context, without interaction with the system.\\
    \textbf{Dependency}             &    37, 38\\
    \textbf{Source}                 &    \cite{Infosec,RichteragileQC}\\
    \textbf{Example}                &    CI/CD pipeline initiates transition when new version is available.\\       
\end{tabular}

\vspace*{2ex}

  \begin{tabular}{p{\mylengthfc}p{\mylength}}
    \textbf{Name}                   &    Context independence\\
    \midrule
    \textbf{ID}                     &    41 \\
    \textbf{Category}               &    System property\\
    \textbf{Description}            &    The requirements and techniques used to implement crypto-agility can also be used for other scenarios.\\
    \textbf{Problem}                &    The approaches used for crypto-agility should not only be applicable in a specific context.\\
    \textbf{Acceptance}             &    The cryptographic approach used must be applicable to at least one other domain.\\
    \textbf{Dependency}             &    20, 32\\
    \textbf{Source}                 &    \cite{OttidentifyPQCchallenges,MehrezProperties}\\
    \textbf{Example}                &    The mechanism to achieve crypto-agility is applicable in both medicine and IoT scenarios.\\       
\end{tabular}

\vspace*{2ex}

  \begin{tabular}{p{\mylengthfc}p{\mylength}}
    \textbf{Name}                   &    Scalability\\
    \midrule
    \textbf{ID}                     &    42 \\
    \textbf{Category}               &    Process\\
    \textbf{Description}            &    The \ca{} implementation can be deployed in additional systems.\\
    \textbf{Problem}                &    If an implementation is always tied to specific platforms, each additional system that is to be made cryptographically agile again means a major development effort.\\
    \textbf{Acceptance}             &    The \ca{} module can be deployed in other systems without in other systems without extensive adaptation.\\
    \textbf{Dependency}             &    20, 32, 41, 44\\
    \textbf{Source}                 &    \cite{Infosec}\\
    \textbf{Example}                &    Wrapper of the \ca{} module for other programming languages exist.\\       
\end{tabular}

\vspace*{2ex}

  \begin{tabular}{p{\mylengthfc}p{\mylength}}
    \textbf{Name}                   &    Real-time\\
    \midrule
    \textbf{ID}                     &    43 \\
    \textbf{Category}               &    Process\\
    \textbf{Description}            &    Modifications to cryptographic functions become active in the production system as immediately as possible.\\
    \textbf{Problem}                &    Modifications to cryptography should not go live only after a long period of time.\\
    \textbf{Acceptance}             &    Added cryptography implementations are reliably ready for use within a defined time period.\\
    \textbf{Dependency}             &    37, 40, 42\\
    \textbf{Source}                 &    \cite{Infosec,NAP24636}\\
    \textbf{Example}                &    Cryptography is outsourced as an external, redundant component so that cryptographic tasks can be dynamically forwarded to different implementations.\\       
\end{tabular}

\vspace*{2ex}

\end{table*}

\begin{table*}
  \caption{Requirements on Level 4 (continued)}

  \begin{tabular}{p{\mylengthfc}p{\mylength}}
    \textbf{Name}                   &    Interoperability\\
    \midrule
    \textbf{ID}                     &    44 \\
    \textbf{Category}               &    System property\\
    \textbf{Description}            &    Different cryptographically agile systems are interoperable with each other.\\
    \textbf{Problem}                &    The compatibility of the cryptography of the global IT infrastructure should not be affected by the introduction of \ca{}.\\
    \textbf{Acceptance}             &    The crypto-agile system enables information exchange with all foreseen communication partners.\\
    \textbf{Dependency}             &    22, 23\\
    \textbf{Source}                 &    \cite{MehrezProperties,housleyRFC7696,NAP24636}\\
    \textbf{Example}                &    A generally accepted crypto-agility specification is introduced, due to which all systems that follow it are interoperable with each other.\\       
\end{tabular}

\end{table*}

}
{ 
}

\end{document}